\documentclass[twocolumn,aps,prb,showpacs,superscriptaddress]{revtex4}
\usepackage[dvips]{graphicx}
\usepackage{bm}
\def\sqr#1#2{{\vcenter{\vbox{\hrule height.#2pt
\hbox{\vrule width.#2pt height#1pt \kern#1pt \vrule width.#2pt}
\hrule height.#2pt}}}}
\def\square{\mathchoice\sqr34\sqr34\sqr{4.2}3\sqr{3.0}3}
\begin{document}
\title{The c-axis transport in naturally-grown
Bi$_2$Sr$_2$CaCu$_2$O$_{8+\delta}$ cross-whisker junctions}
\author{Yu. I. Latyshev}
\email{lat@cplire.ru, yurilatyshev@yahoo.com}
\author{A. P. Orlov}
\author{A. M. Nikitina}\affiliation{Institute of Radio-Engineering and Electronics RAS,
 Mokhavaya 11-7, 101999, Moscow, Russia}
 \author{P. Monceau}
 \affiliation{Centre des Recherches sur les Tr{\`e}s Basses
 Temp{\'e}ratures CNRS, Grenoble 38042, BP 166 Cedex 9, France}
\author{R. A. Klemm}
\email{richard.klemm@und.nodak.edu} \affiliation{Department of
Physics, University of North Dakota, Grand Forks, ND 58202-7129
USA}
\date{\today}
\begin{abstract}
We studied the $c$-axis transport of Bi$_2$Sr$_2$CaCu$_2$O$_{8+\delta}$ (Bi2212) cross-whisker junctions formed by annealing ``naturally''
formed whisker crosses.  These frequently appear during growth when the $ab$-faces of neighboring whiskers come in contact.  We obtained
Fraunhofer patterns of the cross-junction critical currents in a parallel magnetic field, and found a sharp increase in the quasiparticle
tunneling conductance at $eV=50-60$ mV, indicating high junction quality.  For our weak junctions, the interface critical current density is
about 3\% of the critical current density across the stack of bulk intrinsic junctions, as is the room temperature conductivity, and both are
independent of the twist angle, in contrast to most of the data reported on ``artificial'' cross-whisker junctions [Y. Takano {\it et al.},
Phys. Rev. B {\bf 65}, 140513(B) (2002)].  As a minimum, our results provide strong evidence for incoherent  tunneling at least at the
interface, and for at least a small $s$-wave order parameter component in the bulk of Bi2212 for $T\le T_c$.  They are also consistent with the
bicrystal twist experiments of Li {\it et al.} [Phys. Rev. Lett. {\bf 83}, 4160 (1999)].
\end{abstract}
 \vskip0pt\vskip0pt
\maketitle

\section{Introduction}
Since the discovery of the first high temperature superconducting compounds (HTSC),\cite{BM} there has been a huge amount of activity to
understand why this occurs. Although there is nearly universal agreement that the superconductivity arises from the spin-singlet pairing of
holes (or possibly electrons, in some cases), there is no agreement as to the mechanism for this pairing. It is also agreed by nearly all
workers that the most likely place in the HTSC for this pairing to take place is in the ubiquitous CuO$_2$ layers.  Although many proposed
pairing mechanisms varying widely in exoticity (and correspondingly inversely in likelihood) have been suggested, it has so far been exceedingly
difficult to eliminate many of them, and it has been especially difficult to obtain incontrovertible evidence in support of only a single
possible mechanism. In essence, the wide variety of proposed mechanisms falls into two classes: those in which the pairing interaction is
attractive or repulsive in sign. Although there is still no definite method to distinguish these pairing interaction signs, at least it has
generally been agreed that important information in this regard can be obtained if there were to be a strong consensus as to the orbital
symmetry of the superconducting order parameter (OP). Since the superconducting coherence length $\xi$ is comparable to a few lattice constants
at low temperatures $T$, one would further expect the orbital symmetry of the OP to reflect the underlying point group symmetry of the CuO$_2$
planes.\cite{KRS1}

For the tetragonal point group $C_{4v}$ appropriate for some HTSC containing a single CuO$_2$ layer, the relevant group operations for a spin
singlet superconductor are: (a) reflections about the planes normal to the layers containing the directions along the Cu-O bond directions, (b)
reflections about the planes normal to the layers containing the diagonals bisecting neighboring Cu-O bond directions, and (c) rotations by
90$^{\circ}$ about the $c$-axis.  Based upon  oddness or evenness about these group operations,  there are four OP irreducible representations
of $C_{4v}$, which are denoted $s$, $d_{x^2-y^2}$, $d_{xy}$ and $g_{xy(x^2-y^2)}$, respectively.\cite{KRS1}  For example, if the pairing
interaction were attractive, as in the Bardeen-Cooper-Schrieffer (BCS) model, one would expect the OP to most likely have an orbital symmetry
invariant under all of the crystal point group operations, the $s$-wave OP. Although this OP could be highly anisotropic, and could even change
sign at certain points on the Fermi surface, it necessarily has a non-vanishing Fermi surface average. On the other hand, pairing mechanisms
based upon a repulsive interaction necessarily lead to a vanishing Fermi surface average, and generally lead to an OP that is consistent with
$d_{x^2-y^2}$-wave orbital symmetry, which changes sign on opposite sides of the diagonals between the Cu-O bond directions and under
90$^{\circ}$ rotations about the $c$-axis.  For a tetragonal crystal, the OP must have only one of the four symmetries, except below a second
phase transition at $T_{c2}<T_c$, for which a mixed OP form such as $d_{x^2-y^2}+is$ can occur.

In orthorhombic  YBa$_2$Cu$_2$O$_{7-\delta}$, the $b$ axis parallel to the CuO chain direction is longer than the $a$ axis normal to it, and
hence group operations (b) and (c) no longer apply.  In this case, the point group is the orthorhombic $C_{2v}^1$, for which the only effective
group operation is (a), reflections in the mirror planes normal to the layers and containing either of the Cu-O bond directions.\cite{KRS1}  In
this case, $s$- and $d_{x^2-y^2}$-wave OP's can mix without a second phase transition, as can the $d_{xy}$- and $g_{xy(x^2-y^2)}$-wave OP's,
although the relative weight of each component might depend upon $T$. Although,  Bi$_2$Sr$_2$CaCu$_{8+\delta}$ (Bi2212)  is also orthorhombic,
the $b$ axis containing the orthorhombic distortion and the periodic lattice distortion is along a diagonal bisecting neighboring Cu-O bond
directions, leading to an effective point group $C_{2v}^{13}$, with only the group operation (b) remaining.\cite{KRS1} In this case, the
relative mixed OP forms in the absence of a second phase transition are either mixtures of $s$- and $d_{xy}$-wave OP's, or mixtures of
$d_{x^2-y^2}$- and $g_{xy(x^2-y^2)}$-wave OP's. Hence, to the extent that the crystal is perfect, a mixture of $s$- and $d_{x^2-y^2}$-wave OP's
could only occur as a $d_{x^2-y^2}+is$ OP below a second phase transition at $T_{c2}<T_c$.

For the last decade, there has therefore been a raging debate with regards to this $s$-wave/$d$-wave controversy.\cite{Kirtley,Mueller} However,
as the HTSC exhibit a non-superconducting pseudogap in addition to this OP,\cite{pseudogap} many experiments cannot distinguish them very well,
complicating the analysis.  In particular, angle-resolved photoemission spectroscopy (ARPES) and point contact tunneling experiments primarily
measure the quasiparticle density of states, and can infer an overall gap in its spectrum, but cannot infer any information about the phase of
the OP. Although such experiments can infer that both the pseudogap and the superconducting gap arising from the non-vanishing OP below the
superconducting transition temperature $T_c$ can be highly anisotropic, they cannot distinguish if the combined superconducting gap and
pseudogap actually vanishes at some positions in the first Brillouin zone, or is just less than the experimental resolution there, and they
certainly cannot provide any information as to whether it might change sign there.  However, phase-sensitive experiments based upon Josephson
junctions are not affected by the pseudogap, and can distinguish a $d_{x^2-y^2}$-wave OP from a highly anisotropic $s$-wave OP form, such as an
``extended-$s$''-wave OP.  Recently, scanning tunneling microscopy (STM) with atomic resolution have examined surfaces of Bi2212 cleaved at low
$T$, and determined that a disordered array of pseudogap and superconducting regions of characteristic size $2\xi\approx$ 3 nm is stable for
long times.\cite{lang}  If true, this would suggest that there might not be any preferred underlying symmetry relevant for the OP, so that a
mixture of all four OP forms would be possible below $T_c$.  Furthermore, this observation would lend strong support to the notion that the
$c$-axis tunneling across the intrinsic layers in Bi2212 ought to be strongly incoherent.

$c$-Axis bicrystal twist junctions \cite{li} and more recently artificial cross-whisker junctions (CWJ's) \cite{takano1} have attracted
considerable attention because of the possibility of providing phase-sensitive tests of the orbital symmetry of the order parameter (OP) in
(Bi2212).\cite{klemm1,mh,klemm2} With incoherent $c$-axis quasiparticle tunneling, the $c$-axis critical current density $J_c$ across a junction
twisted an angle $\varphi$
 is a constant for $s$-wave or $\propto|\cos(2\varphi)|$ for
 $d$-wave OP's, respectively.\cite{li,bks}  For coherent tunneling, an
 anisotropic Fermi surface causes both OP forms to exhibit a
 strong, four-fold dependence of $J_c(\varphi)$, but a $d$-wave
 OP leads for weak, first-order tunneling to $J_c(45^{\circ})=0$, whereas $J_c(45^{\circ})\ne0$
 for an $s$-wave OP.  The vanishing of $J_c$  for a predominantly $d$-wave OP with weak, first-order coherent tunneling is a
 consequence of the fact that $J_c$ must change sign
 at about 45$^{\circ}$, even in the presence of weak orthorhombicity effects.
 The experimental $J_c(\varphi)$ results are
 still controversial:  in the bicrystal experiments of Li {\it et
 al.},\cite{li} a constant $J_c(\varphi)$ was found, but in the
 artificial CWJ experiments of Takano {\it et al.},\cite{takano1} a strong,
 non-vanishing four-fold $J_c(\varphi)$ was observed.  The quality
 of the Josephson effects on 45$^{\circ}$ CWJ's subsequently
 studied by Takano {\it et al.} was imperfect.\cite{takano2}\\

\section{Bi2212 naturally-grown cross-whisker junctions}

 Here we report on experiments on a new type of twist  junction,
 naturally grown CWJ's.  We found Fraunhofer-like oscillations of
 the critical currents $I_c$ of our CWJ's in parallel magnetic
 fields ${\bf H}$ that clearly indicate dc Josephson behavior across the
 interface thickness $\ell\approx4${\AA}.  This suggests that the
 naturally grown CWJ interface represents a single tunnel junction
 with a small thickness $\ell$.  We also found an increase in the
 quasiparticle tunneling conductivity near
 $eV=2\Delta\approx50-60$ mV that is much sharper than for
 intrinsic stacks of Bi2212 junctions.  We also found $J_c$ to be
 reduced from the bulk value, but independent of $\varphi$.  These
 results  provide strong evidence for the existence of at least a
 small $s$-wave OP component in the bulk of Bi2212.\\

 Bi2212 single crystal whiskers \cite{matsubara} are known to
 possess a high degree of crystalline order.\cite{latyshev1}
 They grow along the $a$-axis direction, independent of the
 crucible or substrate.  The thin whiskers (with thicknesses
 $d<0.3\mu$m and $b$-axis widths $w\le10-20\mu$m) are often free
 of growth steps, macroscopic defects, and dislocations.\cite{latyshev1}  That motivated us to use whiskers to fabricate
 junctions with small twist junction areas.  Takano {\it et al.}
 prepared their CWJ's by placing one whisker upon a MgO substrate,
 a second atop its $ab$ face, and annealing them together.\cite{takano1,takano2}  They reported $J_c(90^{\circ})$ values
 of their CWJ's comparable to the intrinsic $J_c$ of a Bi2212
 junction stack, with a rapid decrease in $J_c(\varphi)$ with
 decreasing $\varphi$, followed by a plateau in $J_c(\varphi)$ for
 $30^{\circ}<\varphi<60^{\circ}$.\cite{takano1,takano2}  However,
 the $I-V$ characteristics of their junctions revealed
 multi-branched structures, suggesting that the interfaces
 themselves consisted of rather ill-defined stacks of about 10
 intrinsic junctions.

\begin{figure}\vskip10pt
{\hskip-60pt\includegraphics[width=0.45\textwidth]{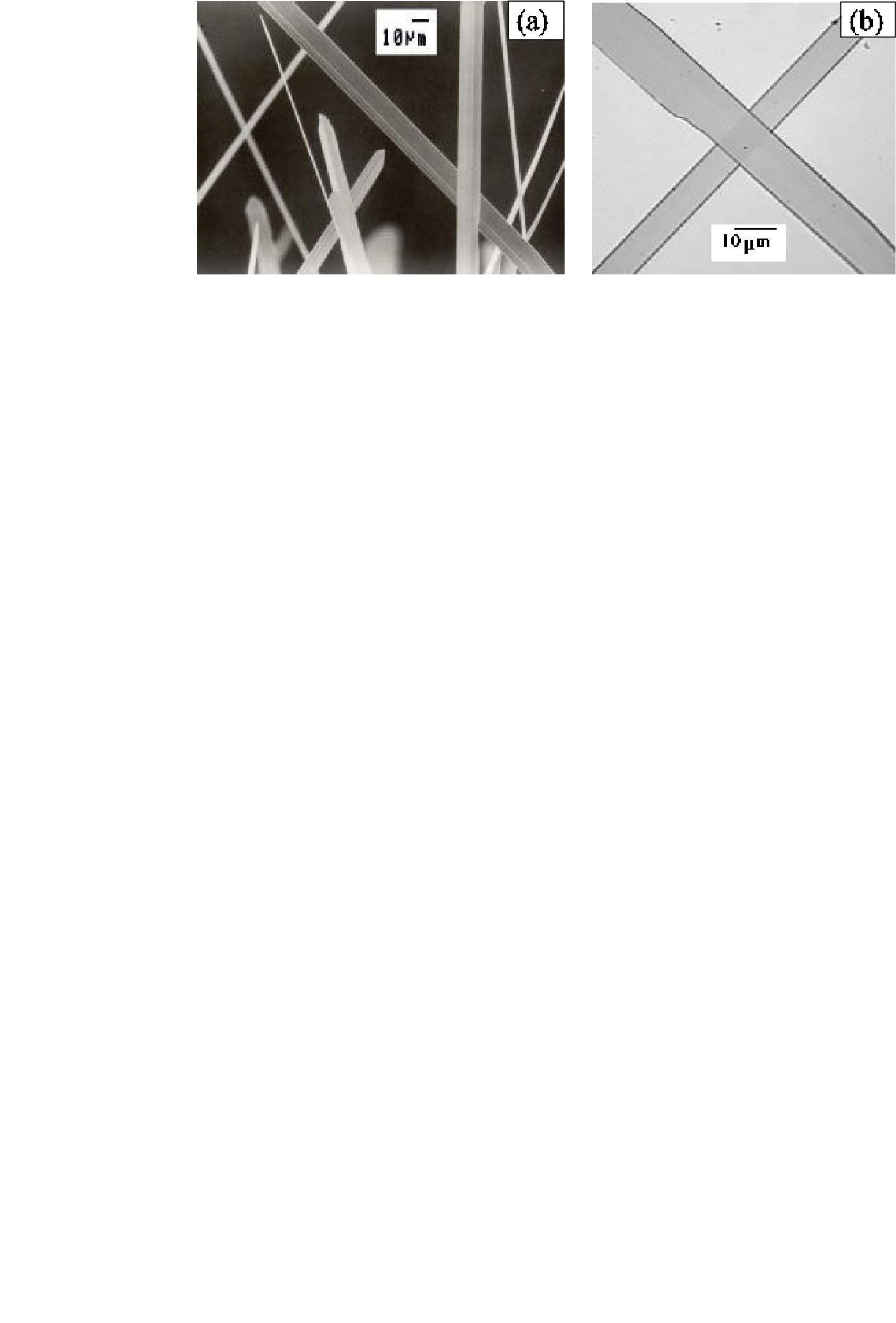}}\vskip-240pt
{\hskip-80pt\includegraphics[width=0.45\textwidth]{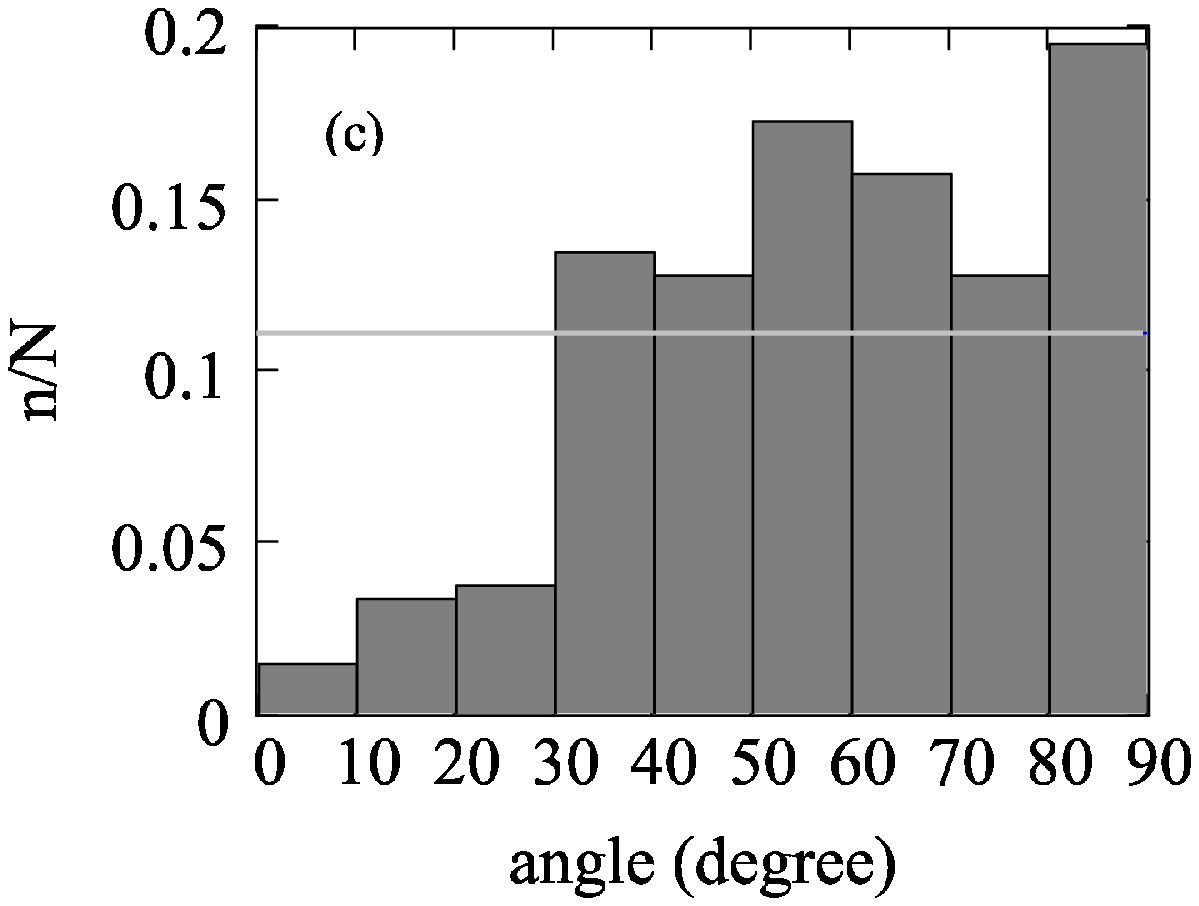}}\vskip-90pt
\caption{SEM pictures of (a) a batch of Bi2212 whiskers containing
natural crosses, (b) of an individual cross-whisker junction, (c)
a histogram of the cross-angle distribution of $N=267$ natural
whisker crosses, where $n/N$ is the relative fraction of the
crosses found within different 10$^{\circ}$ intervals, and the
straight line is the weighted average value.}\label{fig1}
\end{figure}

 In order to obtain CWJ's with more definite interface
 properties, we studied naturally-grown whisker crosses.  Many of
 these form when the $ab$ faces of two whiskers come in contact
 during their growths, \cite{latyshev1} as pictured in Figs. 1a,b.
 The results of an analysis of 267 natural crosses shown in Fig.
 1c reveal a greater abundance of crosses with
 $\varphi>20^{\circ}$, with abundance maxima at 30$^{\circ}$,
 60$^{\circ}$, and 90$^{\circ}$.  However, these as-grown crosses
 have quite high interface resistances $R$ of several tens of
 k$\Omega$ at 300 K, with semiconducting $R(T)$ behavior, and without
 any sign of a superconducting transition temperature $T_c$.
 After annealing in flowing oxygen at $\approx 845^{\circ}$C for
 20 minutes, $R($300K) decreased by 2-3 orders of magnitude and
 the barrier became transparent to a supercurrent below $T_c$.
 Some parameters of seven ``natural'' CWJ's selected for study are
 listed in Table 1.  These cross whiskers were grown in the
 slightly overdoped oxygen stoichiometry regime.\\

\begin{table}
\begin{tabular}{|c|c|c|c|c|c|}
\hline \#&$\varphi$&$S$&$T_{ann}$&$R$&$I_c$\\
&(Deg)&$\mu$m$^2$&$^{\circ}$C&Ohm&$\mu$A\\
\hline 1&56&309&845&33.0&180.0\\2&80&138&845&76.0&62.0\\
3&50&183&847&50.0&200.0\\
4&30&201&845&&80.0\\
5&70&341&840&25.0&\\
6&89&119&840&65.0&172.0\\
7&38&1696&843&7.5&450.0\\
\hline
\end{tabular}
\vskip10pt \caption{Natural cross-whisker junction data.
$\varphi$ is the twist angle, $S$ is the junction area, $T_{ann}$
is the annealing temperature, $R$ is the room temperature
resistance of the cross-whisker junction, and $I_c$ is the
critical current at 4.2 K of the the cross-junction.}\label{tab1}
\end{table}

To demonstrate the electrical uniformity of our fabricated CWJ's,
 a log-log plot of the four-probe interface resistance $R$ at 300
 K of each sample versus the junction area $S$ determined using a
 high resolution optical microscope is presented in Fig. 2.
 Although the $R$, $S$, and $\varphi$ values of our samples varied
 widely, the junctions we tested were all consistent with the
 simple formula $R=R_{\square}/S$ expected for an electrically
 uniform set of junctions with a best fit to the constant
 interface resistance per unit square
 $R_{\square}=10^{-4}\Omega$cm$^2$.  The consistency of this $R_{\square}$ value
 suggests that it is independent of $\varphi$ and that the
 electrical contact area is consistent with the optically
 determined $S$.  For comparison, we estimate the resistance per
 square $R_{\square intr}$ at 300 K for intrinsic $c$-axis
 junctions of single crystal Bi2212 by $\rho_cs$, where $\rho_c$
 is the $c$-axis resistivity and $s$ is the spacing between
 conducting layers.  Using the typical values $\rho_c=10 \Omega$cm
 and $s=1.5$ nm, we estimate $R_{\square
 intr}=1.5\times10^{-6}\Omega$cm$^2\approx R_{\square}/60$.

\begin{figure}\vskip0pt
{\hskip-50pt\includegraphics[width=0.45\textwidth]{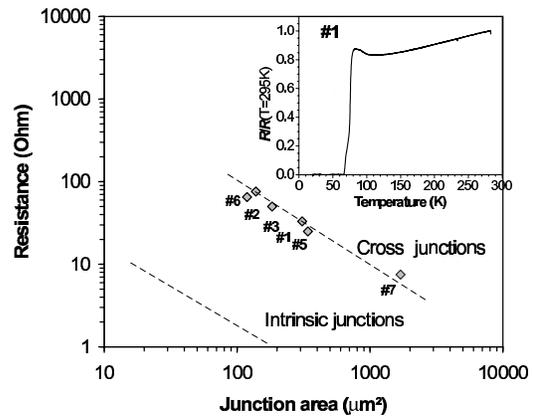}}\vskip-190pt
\caption{The dependence of the cross-junction resistance on its
area for the junctions listed in Table 1.  The inset shows $R(T)$
for a natural cross-whisker junction.}\label{fig2}
\end{figure}

\section{Superconducting results}

A typical $R(T)$ for a CWJ measured with an ac current $\sim 1\mu$A is shown in the inset of Fig. 2.  This $T$ dependence is typical of that for
$\rho_c(T)$ for slightly overdoped single crystals.\cite{watanabe}  Below $T_c$, the low-$T$ $I-V$ characteristics of CWJ's pictured in Fig. 3a
show a long, linear $I(V)$ quasiparticle branch region of the tunneling type at low bias voltages $V$, followed by a sharp rise in $I$ at
$V_g=50-60$ mV, accompanied by a switch to the normal, resistive state at some current $I_{sw}$.  The current density corresponding to this
switch $J_{sw}=I_{sw}/S$ was found to be $\sim2$kA/cm$^2$ for three CWJ's studied at high currents.  This $J_{sw}$ value corresponds to $J_c$
for the intrinsic junctions in slightly overdoped Bi2212 stacks,\cite{latyshev2,inomata} suggesting that the switching may be associated with
the spreading of the resistive state inside the bulk of the whiskers in contact.  In the subgap bias region $V<V_g$, the quasiparticle branch
exhibits fine structure characterized by 10-20 jumps in $V$ with increasing $I$ which are 1-2 mV in magnitude.  Application of an $H$ of several
T parallel to the layers removes these $V$ jumps, as shown in Fig. 3b.  More details of this fine structure will be published
elsewhere.\\

\begin{figure}\vskip0pt
{\hskip-110pt\includegraphics[width=0.45\textwidth]{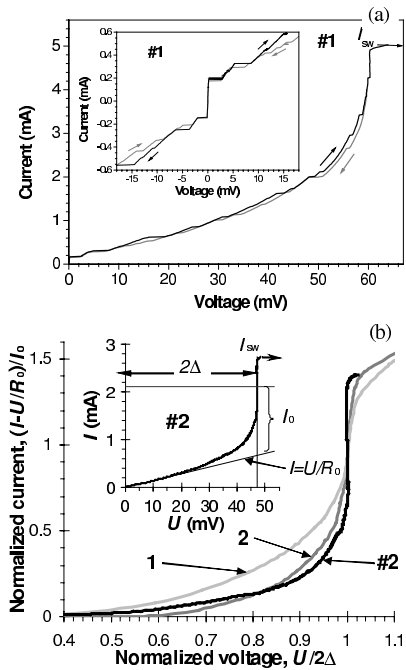}}\vskip-90pt
\caption{The $I-V$ characteristics of Bi2212 cross-whisker
junctions: (a) for junction \#1 at large and small voltage scales;
(b) a comparison of the normalized $I-V$ characteristics with the
initial linear parts subtracted of Bi2212 junctions of different
types: an intrinsic junction within a Bi2212 stack  (curve
1),\cite{latyshev1} a Bi2212/Dy2278/Bi2212 junction  (curve
2),\cite{bozovic} and our cross-whisker junction \#2, where $I_c$
and the subgap structure are suppressed by a 4T parallel magnetic
field.  The inset shows the original, unsubtracted $I-V$ curve of
sample \#2 in the same field.
 See text.}\label{fig3}
\end{figure}

One of the most remarkable features of natural CWJ's is the very sharp increase in $I(V)$ at $eV\approx50-60$ mV, pictured in Figs. 3a,b,
accompanied by a vanishing of the dynamic resistance.  This behavior is expected for superconducting-insulating-superconducting (SIS) junctions
at $eV=2\Delta$. Values of $2\Delta=50-60$ mV were obtained both from intrinsic tunneling experiments on slightly overdoped Bi2212 mesas using a
pulsed voltage technique to avoid self-heating effects,\cite{suzuki} and from scanning tunneling microscopy (STM) measurements.\cite{renner}
Hence, our CWJ interfaces are also likely to be elementary single junctions with highly suppressed self-heating effects.  Experiments on mesas
of Bi$_2$Sr$_2$Ca$_2$Cu$_3$O$_{10+\delta}$ containing a single
intrinsic junction led to the same conclusion.\cite{odagawa}\\

In Fig. 3b, we compare our CWJ interface $I/V$ characteristics of sample \#2 with those of two other sample types, plotting the curves with the
linear part of the quasiparticle branch subtracted.  The subtracted quasiparticle branch of the intrinsic stacked junction $I/V$ data, curve 1,
differs considerably from our subtracted CWJ interface data, and shows a much more smooth increase near $eV=2\Delta$.  In curve 2 we show the
earlier data obtained from artificial Bi2212 structures containing a single insulating Bi$_2$Sr$_2$Dy$_x$Ca$_{7-x}$Cu$_8$O$_{20+y}$ (Dy2278)
layer.\cite{bozovic}  In that experiment the $I-V$ characteristics also have a subtracted, long linear initial part and a very sharp
quasiparticle current increase at $V_g=50$ mV, corresponding to $eV=2\Delta$.

Studies of $I_c$ in a parallel magnetic field ${\bf H}$ show
Fraunhofer-like oscillations,
\begin{eqnarray}
I_c&=&I_{c0}\Bigl|\frac{\sin x}{x}\Bigr|+I_{c1},
\end{eqnarray}
where $x=\pi Hw\ell/\Phi_0$, $\Phi_0$ is the flux quantum, and $w$ and $\ell$ are the in-plane junction width $\perp{\bf H}$ and effective
junction thickness, respectively.  The non-oscillation background part of $I_c(H)$, $I_{c1}$, was only 10\% of $I_c(0)$ for the best junction,
\#2, but can be as large as 50\% of $I_c(0)$ for a lower quality junction such as \#6, as shown in Fig. 4.  In the intrinsic stacked junction
$\ell=c/2=15.6${\AA}, one-half the $c$-axis lattice parameter.\cite{latyshev2} Remarkably, for natural CWJ's we reproducibly found
$\ell\approx4${\AA}, about 4 times smaller than $c/2$.  That indicates that the CWJ interfaces contain only one insulating layer.  It is well
known that a regular Bi2212 crystalline structure contains two insulating distances between elementary conducting layers.  One is a short
distance of 3{\AA} formed by the Ca layer between single CuO$_2$ layers.  The larger distance of 15.6{\AA} is associated with the coupling of
the cuprate bilayers.\cite{latyshev2}  That contains two BiO and two SrO layers.  If the only parameter involved in the tunneling matrix
elements were the junction thickness, then one might infer that the CWJ interface were related to the shorter elementary CuO$_2$ interlayer
distance, and that the interfaces would be the terminating layer of each contacting whisker.  However, since the interfaces lead to a factor 60
weaker transparency in the normal state than do the intrinsic junctions, a more likely scenario is that the interfaces uniformly contain some
more
strongly insulating oxide barrier.\\

\begin{figure}\vskip0pt
{\hskip-70pt\includegraphics[width=0.45\textwidth]{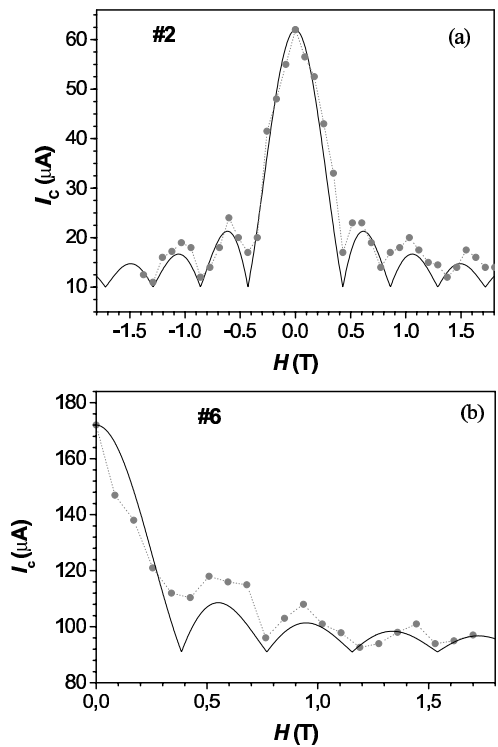}}\vskip-60pt
\caption{Dependence of the critical current $I_c$ of
cross-junctions \#2 (a) and \#6 (b) on the parallel magnetic field
$H$ at $T=4.2$ K.  The solid lines correspond to the function
$I_c=I_{c0}|\sin x/x|+I_{c1}$ with $x=\pi H w\ell/\Phi_0$, where
$w$ is the width of the junction and $\ell$ is the thickness of
the junction.  For \#2 with parameters $\Delta H=0.39$ T, $w=13.6
\mu$m, $\ell=3.9${\AA}, $I_{c0}/I_c=0.47$.}\label{fig4}
\end{figure}

To test the $J_c(\varphi)$ dependence, we measured $J_c$ for 6 natural CWJ's with various twist angles $\varphi$ (see Table 1). Our data differ
significantly from those found for ``artificial'' crosses studied by Takano {\it et al.},\cite{takano1,takano2} as shown in Fig. 5.  We defined
$J_c$ in three different ways: (1) from $J_c=I_c/S$ (stars), (2) from $J_c=I_{co}/S$, where $I_{c0}$ is the amplitude of the oscillating
$I_c(H)$ defined in Eq. (1) (filled circles), and (3) from the switching current $I_{sw}$ into the resistive state (filled squares).  If
$I_{sw}$ corresponds to the bulk intrinsic $J_{cb}\approx2\times10^3$ A/cm$^2${},\cite{latyshev2,inomata} then $J_c=(I_c/I_{sw})J_{cb}$. As seen
from Fig. 5, the values of $J_c$ defined in these three different ways are roughly consistent with each other for each sample.  The most
reliable data obtained by the second method were available only for samples \#2 and \#6.  However, averaging all of the data for 6 samples shows
a $\varphi$-independent $J_c$ (dashed line) with the averaged $J_c\approx50$A/cm$^2$, a factor 20-40 smaller than $J_c$ for bulk intrinsic
junctions.

\begin{figure}\vskip0pt
{\hskip-100pt\includegraphics[width=0.45\textwidth]{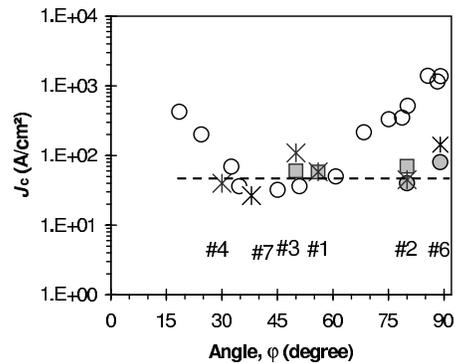}}\vskip-250pt
\caption{Dependencies of the critical current density $J_c$ of
Bi2212 natural cross-whisker junctions on the twist angle
$\varphi$.  The full symbols and stars correspond to
naturally-grown cross-whisker junctions, the different symbols
relating to the three different definitions of $J_c$ given in the
text.
 The open circles correspond to the data of Takano {\it et al.}
  for ``artificial'' cross-whisker junctions.\cite{takano2}}\label{fig5}
\end{figure}

An angularly independent $J_c(\varphi)$ data set was reported for bicrystal twist junctions.\cite{li}  Those authors  measured the same $J_c$ as
the interface as in the bulk, and attributed this result to $s$-wave symmetry of the OP in the crystal bulk.\cite{li} The experiment was done,
however, on large crystals with an in-plane cross section 100-300$\mu$m and with a somewhat reduced critical current density
$J_c\sim200$A/cm$^2$ at low $T$.\cite{li}  They did not present convincing evidence for Josephson behavior of the interface.  However, as they
observed the same $J_c$ values for their twist junctions as for their single crystal junctions at 0.9$T_c$, their low-$T$ twist junction $J_c$
values
were much larger than our CWJ interface $J_c$ values.\\

The experiments of Takano {\it et al.}\cite{takano1,takano2} on artificial CWJ's showed high $J_c$ values of 1.5$\times10^3$A/cm$^2$ at
$\varphi=90^{\circ}$ that decreased with decreasing $\varphi$, exhibiting an extended, flat minimum for $30^{\circ}<\varphi<60^{\circ}$.  They
initially considered that strong $\varphi$ dependence of $J_c(\varphi)$ to be evidence for a predominant $d$-wave OP symmetry.  However, they
subsequently found reproducible (non-vanishing) $J_c(45^{\circ})$ values in many artificial CWJ's, and for these $\varphi\approx45^{\circ}$
junctions they found Fraunhofer-like $I_c(H)$ patterns with a high background value of $\approx$50\% of $I_c(0)$.\cite{takano2}  For other
angles the backgrounds of the Josephson $J_c$ values were not analyzed in this way. Very recently, they also presented Shapiro step data on a
45$^{\circ}$ artificial CWJ.\cite{takano3}  Combined with the Fraunhofer data, the Shapiro step analysis provided strong evidence that their
artificial CWJ's contained only weak, first-order quasiparticle CWJ tunneling.  They also showed that the strong $J_c(\varphi)$ they obtained
for their artificial CWJ's was independent of $T$ for 5K$\le T\le 60$K.  Thus, they concluded that the superconducting gap did not vanish in the
bulk, even along the ``nodal direction'', for
temperatures up to at least 60K.\cite{takano3} \\

For our natural CWJ's, we found an angularly-independent $J_c$ of about 50A/cm$^2$, 30 times smaller than the Takano {\it et al.} data for
$\varphi=90^{\circ}$, but very close to their data for $30^{\circ}<\varphi<60^{\circ}$, as shown in Fig. 5.  Our data near $\varphi=90^{\circ}$,
however, were confirmed by Fraunhofer patterns.  Anyway, the $J_c$ values of our junctions are much lower than the intrinsic $J_{cb}$ values
measured on mesas fabricated from the same Bi2212 whiskers.\cite{latyshev2}  The $c$-axis transport and magneto-transport on those mesas at low
temperatures were well described by a $d$-wave Fermi-liquid model with a significant amount of coherent interlayer tunneling.
\cite{latyshev3,morozov}  In that model one would expect to observe a strong four-fold $J_c(\varphi)$ dependence at the interface, which {\it
vanished} at $45^{\circ}$.\cite{klemm1,krs} As one possible  qualitative explanation of  the reduced and angularly-independent $J_c$ through the
interface of our natural CWJ's, we suggest that the scattering at the interface of the twist junctions might be highly incoherent due to either
the breaking of translational symmetry at the interface, or to junction disorder.  The former could impose a mixed order parameter of the $d+is$
type, with a subdominant $s$-component in the layers near to the interfaces, at least at low $T$ \cite{krs}. However, such behavior is not
expected near to $T_c$.\cite{krs} We therefore measured the temperature dependence of $J_c$ for two natural CWJ's with cross-whisker angles
38$^{\circ}$ and 86$^{\circ}$, and the results are presented in Fig. 6.  We conclude that there is no qualitative difference in the onset of
$J_c$ for these two $\varphi$ values, arguing strongly against that $d+is$ scenario, as any $s$-wave component would have to be present at 68K.
In addition, translational symmetry breaking would cause the quasiparticles to change their momentum locally in tunneling from one atomic site
to another one on the opposite side of the junction, which would be displaced parallel to the junction in real space even for a 90$^{\circ}$
junction. However, the quasiparticles on each side of the junction have a well-defined wave vectors ${\bm k}$ and ${\bm k}'$, respectively, and
the ones most likely to contribute to the tunneling have  dispersions $\xi({\bm k})=\epsilon({\bm k})-E_F$ and $\xi({\bm k}')$ that are small on
both sides of the junction. As shown for $c$-axis twist junctions,\cite{arnold,bks} for bandwidths consistent with ARPES experiments, it is
still possible to have quasielastic coherent tunneling that is only weakly suppressed from that for intrinsic, untwisted junctions for twist
angles up to 2-5$^{\circ}$, regardless of the OP symmetry.\cite{bks} Hence, interface imperfections pose a more likely origin for any possible
incoherent interface tunneling. However, we do not have any specific experimental evidence to demonstrate conclusively that the interfaces are
disordered. In the absence of any such evidence, we have to also consider the possibility that the tunneling between the intrinsic layers of
Bi2212 might also be incoherent, consistent with the STM observations
of Lang {\it et al.}\cite{lang}\\

\begin{figure}
\includegraphics[width=0.45\textwidth]{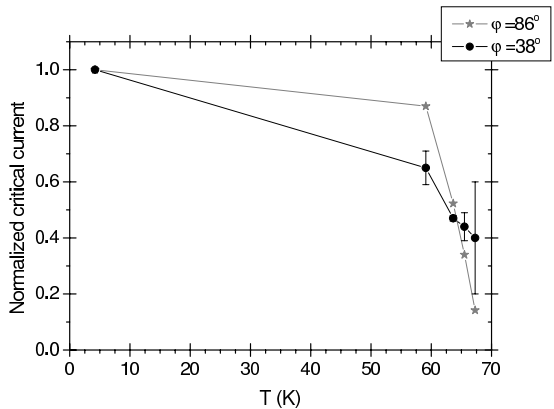}
\caption{$J_c(T)/J_c(4.2$K) for two natural CWJ's with $\varphi=38^{\circ}, 86^{\circ}$. }\label{fig6}\end{figure}

The $R_{\square}$ of our twist junctions is a factor of 60 higher than for the individual intrinsic junctions, in spite of the lower effective
barrier thicknesses.  However, we note that $J_c$ is only a  factor 20-40 lower than for the bulk intrinsic junctions, suggesting that if $R$
were to represent the intrinsic, low-$T$ values of $R_n$, $I_cR_n$ for our CWJ's would be at least as large as those for intrinsic Bi2212 single
crystal junctions.  Thus, it seems that a likely explanation for our values of $J_c$ being lower than those obtained from intrinsic bulk
junctions is simply due to the weaker tunneling matrix elements, as evidenced by the larger $R$ values of our CWJ's.   Because of the presumed
strongly incoherent scattering at the interface, any $d$-wave component to the critical current would be completely suppressed, and the observed
small critical current  might be due to the remaining $s$-wave bulk component.  This qualitative explanation implies a reduced $T_c$ value of
the junction $T_{cj}$ relative to the bulk value $T_{c0}$.\cite{krs}  In our experiments we observed a reduction of $T_{cj}$ by about 8K below
the intrinsic $T_{c0}$ of the whiskers ($T_{c0}=76$K), which places a lower limit ($T_{cj}=68$K) on the $s$-wave $T_c$ value.  However, this
$T_{cj}$ reduction could arise from our annealing process, as unannealed samples were not superconducting.  In the Li {\it et al.} data,
\cite{li} the reduction in $T_c$ from the twist junctions was only about 1K from the bulk values.  The fact that our data for $J_c$ are close to
the Takano {\it et al.} data \cite{takano2} at $30^{\circ}<\varphi<60^{\circ}$ may be an indication of the presence of an $s$-wave OP component
of the same strength in their cross-whisker junctions as well.  On the other hand, both sets of low $J_c$ values could just be due to similar
$R_{\square}$ values characteristic of similarly weak tunnel barriers, and that the OP was predominantly $s$-wave.  We remark that the presence
of at least a small $s$-wave component of the OP was also reported at the $c$-axis
interface of Bi2212/Pb Josephson junctions.\cite{moessle}\\

We remark that the sharp increase of the quasiparticle conductivity at $eV=2\Delta$ may also be a signature of the presence of a rather isotopic
$s$-wave component of the OP in our junctions and in Bi2212/Dy2278/Bi2212 junctions.\cite{bozovic}  This might suggest that the
superconductivity could arise primarily on the saddle bands near the $\overline{M}$ points in the first Brillouin zone, as suggested by Tachiki
{\it et al.},\cite{tachiki} and would appear to be rather constant for either $s$- or $d$-wave superconductors.  For a substantially $d$-wave OP
with a gap  on the regular Fermi surface at the interface, this onset would
be expected to be very broad.\\

\section{Conclusions}

Experiments on naturally-grown and annealed Bi2212 cross-whisker junctions show a small effective interface thickness $\approx 4${\AA} and a
very sharp quasiparticle gap edge in their $I-V$ characteristics, in contrast to intrinsic Josephson junctions in bulk single crystals.  We also
found that the Josephson critical current density at the interface is significantly reduced from the intrinsic bulk value, and is insensitive to
the twist angle. However, this reduction in the critical current density may simply be a consequence of more comparably greatly increased normal
state resistance at the interface, which is also independent of the twist angle.  As a minimum, we infer incoherent quasiparticle and Josephson
tunneling at least at the interface, and the presence of at least a small (3\% of the total or greater) $s$-wave component of the order
parameter in the bulk of the samples for $T\le T_{cj}=68$K.  Our results on natural CWJ's are also consistent with the the Li {\it et al.}
bicrystal twist experiments.\cite{li}

\section*{Acknowledgments}
We would like to thank L. N. Bulaevskii, Ch. Helm, Y. Takano, T.
Hatano, T. Yamashita, A. Koshelev, I. Bozovi{\'c}, K. Scharnberg,
and N. Pavlenko for fruitful discussions.  We acknowledge support
from CRDF grant No. RPI-12397-MO-02, from grant No.
40.012.1.111.46 from the Russian Ministry of Science and Industry,
and from the Jumelage project No. 03-02-2201 between the IRE RAS
and the CRTBT CNRS.

\end{document}